\documentclass[prl,twocolumn,superscriptaddress,floatfix]{revtex4-1}
\usepackage{color}
\usepackage{amsmath}
\usepackage{amssymb}
\usepackage{bm}      	
\usepackage{graphicx}
\usepackage[unicode=true,pdfusetitle,
 bookmarks=false,bookmarksnumbered=false,bookmarksopen=false,
 breaklinks=false,pdfborder={0 0 1},backref=false,colorlinks=true,urlcolor=blue,citecolor=blue,linkcolor=black]
 {hyperref}
\usepackage{subfigure}
\usepackage{textcomp}

\newcommand{\bsub}{\begin{subequations}}
\newcommand{\esub}{\end{subequations}}
\newcommand{\tel}{\tau_{\mathsf{el}}}
\newcommand{\vex}[1]{\bm{\mathrm{#1}}}

\begin{document}
\title{Enhanced Amplitude for Superconductivity
due to Spectrum-wide Wave Function Criticality 
in Quasiperiodic and Power-law Random Hopping Models}
\def\rice{Department of Physics and Astronomy, Rice University, Houston, Texas
77005, USA}
\def\rcqm{Rice Center for Quantum Materials, Rice University, Houston, Texas
77005, USA}
\author{Xinghai Zhang}\affiliation{\rice}
\author{Matthew S.\ Foster}\affiliation{\rice}\affiliation{\rcqm}
\date{\today}

\begin{abstract}
We study the interplay of superconductivity and a wide spectrum of critical (multifractal) wave functions 
(``spectrum-wide quantum criticality,'' SWQC) in the one-dimensional Aubry-Andr\'{e} and power-law 
random-banded matrix models with attractive interactions, using self-consistent BCS theory. We find that 
SWQC survives the incorporation of attractive interactions at the Anderson localization transition,
while the pairing amplitude is maximized near this transition in both models. Our results suggest that SWQC, 
recently discovered in two-dimensional topological surface-state and nodal superconductor models, 
can robustly enhance superconductivity. 
\end{abstract}
\maketitle

Bulk low-temperature superconductors typically reside in the so-called ``dirty limit'' $\Delta \ll 1/\tel$, 
where $\Delta$ is the spatially averaged order parameter amplitude and $1/\tel$ is the elastic scattering rate. As long
as the normal state is a good conductor ($\varepsilon_F \tel \gg 1$, where $\varepsilon_F$ is the Fermi energy),
non-magnetic disorder has a negligible effect on $T_c$ (Anderson's theorem \cite{AG1959, Anderson1959}).
Unconventional superconductors like the cuprates and twisted-bilayer graphene \cite{Cao2018,Andrei2020}
are effectively two-dimensional (2D), however, where arbitrarily weak disorder typically induces Anderson 
localization of all electronic states \cite{Lee1985}. 

The competition between disorder and superconductivity is responsible for the superconductor-insulator transition 
\cite{Ma1985}, which has been a subject of extensive study (see e.g.\ Refs.~\cite{Feigelman2010a,Trivedi2012,Goldman2012,Burmistrov2015}). 
Self-consistent numerical solutions to the Bogoliubov-de Gennes (BdG) equations revealed that strong disorder, which localizes single-particle states, 
can induce emergent granularity in $\Delta(\vex{r})$ \cite{Ghosal1998,Ghosal2001,Fan2020}. 
This augments phase fluctuations that ultimately destroy superconductivity 
\cite{Ghosal1998,Ghosal2001,Mayoh2015,Dubi2007,Erez2013}. 

A surprising recent development was the realization that superconductivity can sometimes be \emph{enhanced}
by random or structured inhomogeneity \cite{Arrigoni2004,Martin2005,Kivelson2007,Aryanpour2006,Aryanpour2007,Feigelman2007,Zhou2008,Tsai2008,Mondaini2008,
Feigelman2010,Burmistrov2012,Kravtsov2012a,DellAnna2013,Mondaini2012,Mayoh2015,Dodaro2018,Tezuka2010,Gastiasoro2018,Burmistrov2021}.
In particular, near the bulk Anderson metal-insulator transition or generally for weak disorder in 2D, the critical rarification
(multifractality \cite{Evers2008}) of single-particle wave functions induced by quantum interference can enhance interaction 
matrix elements \cite{Feigelman2007,Feigelman2010,Burmistrov2012,Foster2012,Foster2014}.
The multifractal wave functions have larger spatial overlap and stronger state-to-state correlations 
for states with similar energies (``Chalker scaling'' \cite{Chalker1988,Chalker1990,Cuevas2007}),
and therefore interaction effects are stronger compared to that for extended or localized ones.
It was argued that this can boost both the superconducting order parameter amplitude $\Delta$ and $T_c$ \cite{Feigelman2007,Feigelman2010,Burmistrov2012,Mayoh2015,Fan2020,Fan2020a,Fan2021,Stosiek2021}.
Multifractal order parameter modulations have recently been observed in experiments on 2D superconductors 
\cite{Zhao2019, Rubio2020, Carbillet2020, Siebarski2020, Siebarski2021}.

In this paper, we consider a new twist on this theme. In particular, we show that the superconducting amplitude can
be strongly enhanced for a system with a wide spectrum of multifractal single-particle wave functions,
a phenomenon dubbed ``spectrum-wide quantum criticality'' (SWQC). SWQC was very recently discovered
to arise robustly in 2D surface-state theories with disorder \cite{Ghorashi2018,Sbierski2020,Ghorashi2020,Karcher2021}. 
These theories describe surface states of model bulk topological superconductors \cite{Foster2014}, as well 
as nodal quasiparticles in dirty 2D $d$-wave superconductors \cite{Altland2002,Ghorashi2020}.
In these theories, SWQC may be protected by a robust topological mechanism \cite{Karcher2021}.

\begin{figure}[t!]
  \centering
  \includegraphics[width=0.48\textwidth]{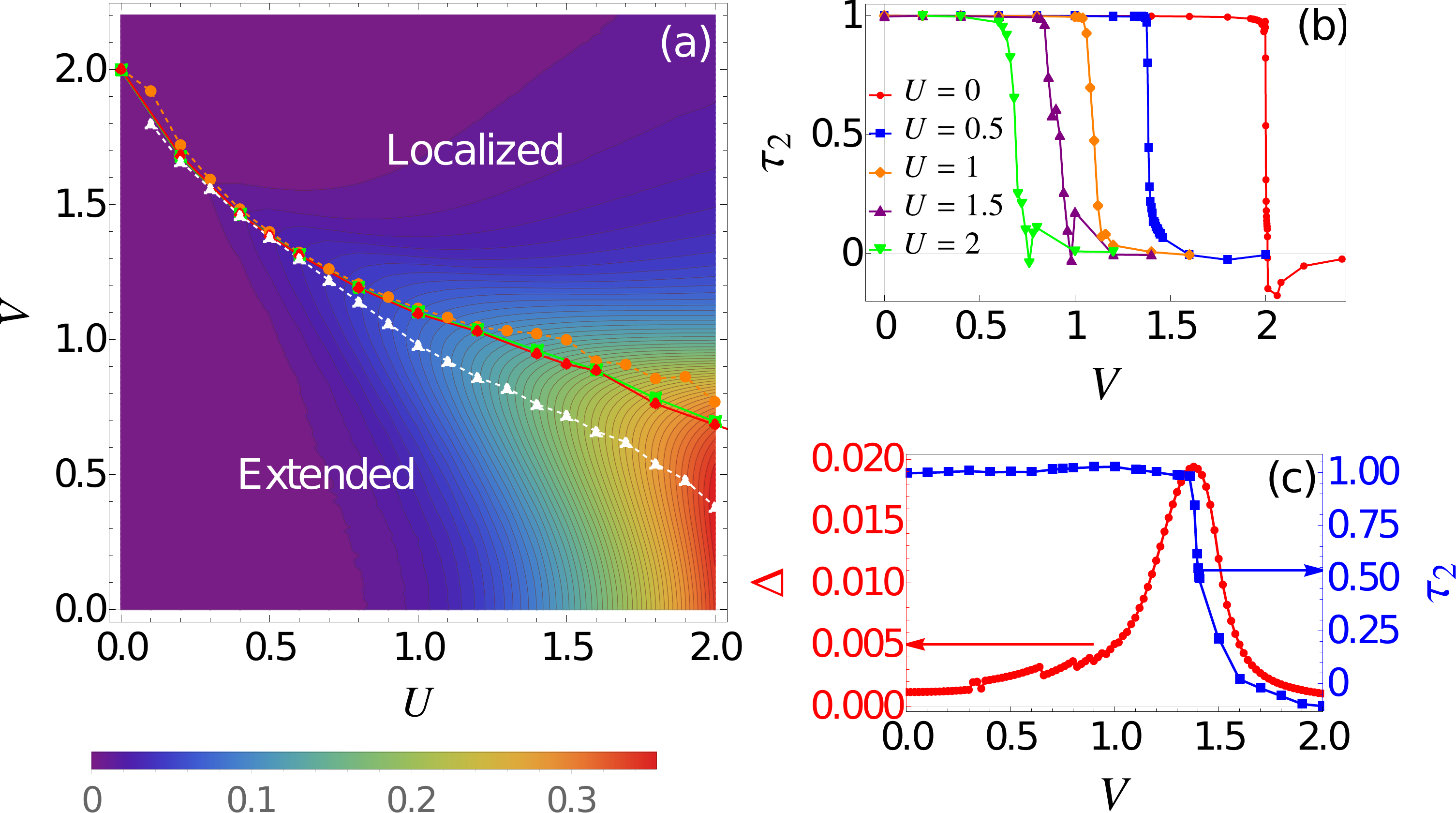}
  \caption{The enhancement of superconductivity in the Aubry-Andr\'{e} (AA) model with attractive Hubbard interactions, Eq.~\eqref{eq:H-AA} with $t=1$. 
	(a): Contour plot of $\Delta$ versus attractive interaction strength $U$ and incommensurate potential strength $V$. 
	The orange ($L_1=1597$, $L_2=2584$), green ($L_1=2584$, $L_2=4181$) and red ($L_1=4181$, $L_2=6765$) curves represent the MIT obtained from 
	scaling of second multifractal dimensions with different system sizes. The strongest enhanced superconductivity with fixed interaction is indicated by the white curve. 
	(b): $\tau_2$-$V$ for different interaction strengths. 
	$\tau_2$ is obtained from the spectrum-averaged inverse participation ratio $\left<P_q\right>\sim L^{-\tau_q}$ with two system sizes $L_1=4181$ and $L_2=6765$. 
	The Anderson transition occurs near the sharp drop of $\tau_2$.
	(c): $\Delta$ and $\tau_2$ versus $V$ with $U=0.5$. $\Delta$ peaks around the MIT, where $\tau_2$ drops sharply.}
	\label{fig:DeltaUV-AA}
\end{figure}

In this work we perform numerical self-consistent BdG calculations on special 1D systems also known to exhibit SWQC,
when fine-tuned to the Anderson metal-insulator transition (MIT). (Working in 1D permits us to access much larger
system sizes than would be possible in 2D). In particular, we consider the effect of attractive Hubbard
interactions for spin-1/2 fermions in the quasiperiodic Aubre-Andr\'e and power-law random-banded matrix models. 
Quasiperiodic systems have recently garnered a surge of interest due to realizations with ultracold atoms 
\cite{Guidoni1997, Roati2008, Palencia2010, Deissler2010, Schreiber2015, Bordia2016, Bordia2017, Viebahn2019, Szabo2020, Sbroscia2020, Gautier2021},
applications in many-body localization physics 
\cite{Iyer2013,Agarwal2017,Abanin2019,Schreiber2015,Bordia2016,Bordia2017,Sbroscia2020},
Hofstadter superconductivity \cite{Shaffer2021, Shaffer2022}, 
and
progress in moir\'{e} materials \cite{Bistritzer2011, Santos2007, Cao2018, Cao2018a, Andrei2020} 
with large twist angles \cite{Moon2013, Spurrier2019, Pal2019, Mirzakhani2020, Bucko2021, Yao2018, Ahn2018, Bocquet2020}.
The Aubry-Andr\'{e} (AA) model \cite{Aubry1980, Sokoloff1985} is a canonical example of a 1D quasiperiodic system. While its energy spectrum is well-known to 
possess fractal structure (the Hofstadter butterfly \cite{Hofstadter1976,Tang1986}), a less-appreciated aspect is the fractality of 
the corresponding wave functions, which exhibit SWQC at the MIT tuned by the incommensurate potential strength \cite{Evangelou1987,Siebesma1987}.
SWQC also occurs in the ensemble of power-law random banded matrices (PRBM) 
\cite{Mirlin2000a,Evers2008,Mirlin1996,Kravtsov1997,Evers2000,Mirlin2000,Cuevas2007,Kravtsov2010,Kravtsov2012,Kravtsov2015}.

\begin{figure}[t]
  \centering
  \includegraphics[width=0.48\textwidth]{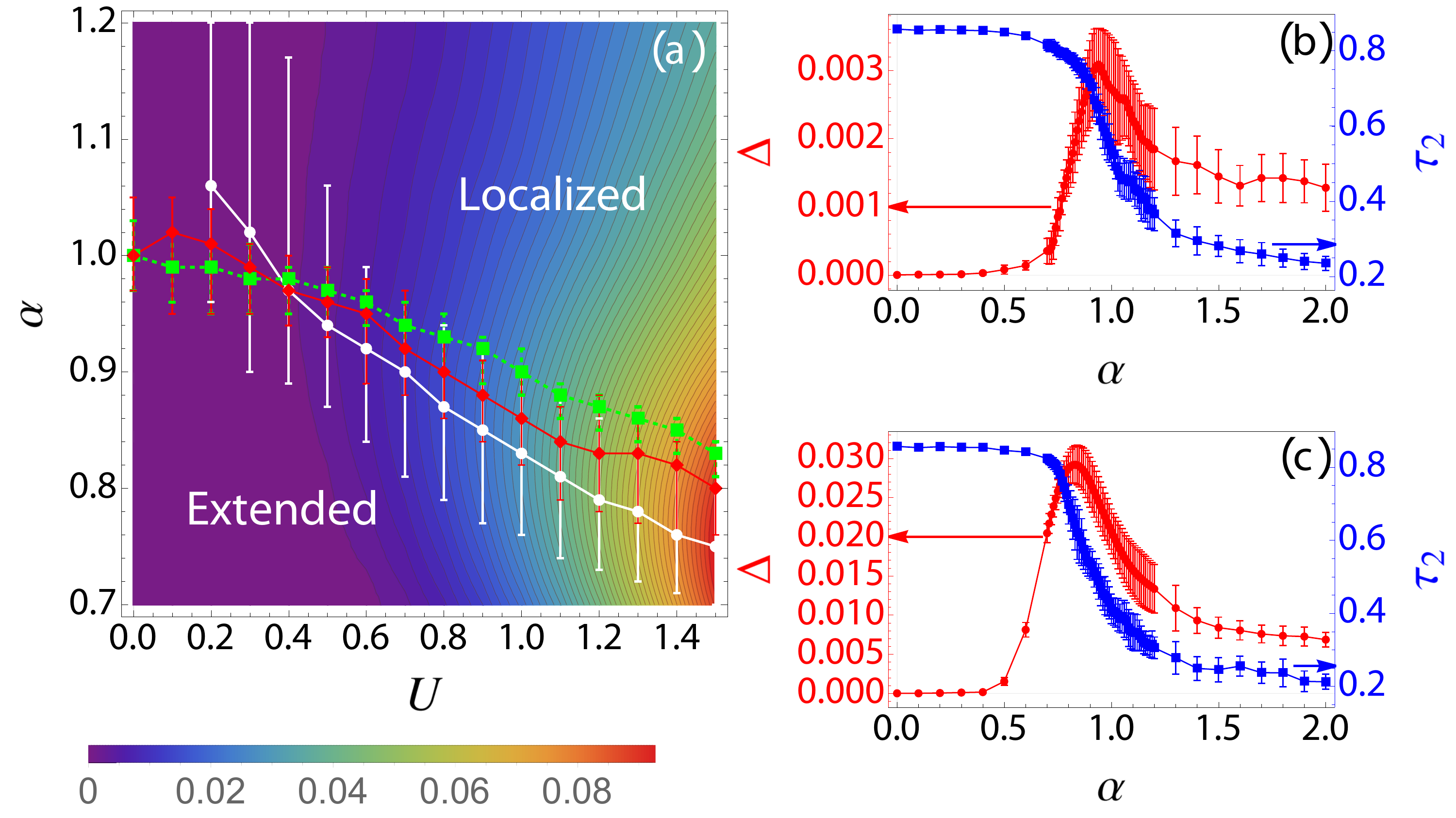}
  \caption{The enhancement of superconductivity in the PRBM model with attractive Hubbard interactions, Eq.~\eqref{eq:HPRBM}. 
	(a): Contour plot of $ \Delta$ versus $U$ and the decay power $\alpha$, where $\left<\cdots\right>$ stands for 
	disorder averaging of $20$ samples of size $L=2000$ systems. 
	The MIT occurs at $\alpha=1$ without interactions, and we use the value of 
	$\tau_2=-\left<\log(P_2)/\log(L)\right>$ 
	at $U=0$ and $\alpha=1$ as the criteria for the MIT at for nonzero $U$. 
	The red and green curves represent the MIT obtained by fitting $\tau_2$ using the lowest-lying quasiparticle state and the average of $1\%$ 
	of the low-lying states, respectively. The most enhanced superconductivity is indicated by the white curve. 
	The error bars are obtained from the standard deviation of $\Delta$ and $\tau_2$ due to disorder averaging, converted to uncertainty in $\alpha$. 
	The large error bars reflect the broad enhancement region of superconductivity. 
	(b), (c): $\Delta$ and $\tau_2$ as functions of $\alpha$ for $U=0.5$ and $U=1$, respectively. 
	$\Delta$ peaks around the MIT, indicating multifractal enhancement of the order parameter. 
	The wide region of wave function criticality (indicated by the rather slower decrease of $\tau_2$ compared to the case of AA model) 
	explains why the maximal enhancement curve [white in (a)] does not follow MIT curve as well as the case of AA model.}
  \label{fig:Delta-PRBM}
\end{figure}

\begin{figure}[t]
  \centering
  \includegraphics[width=0.48\textwidth]{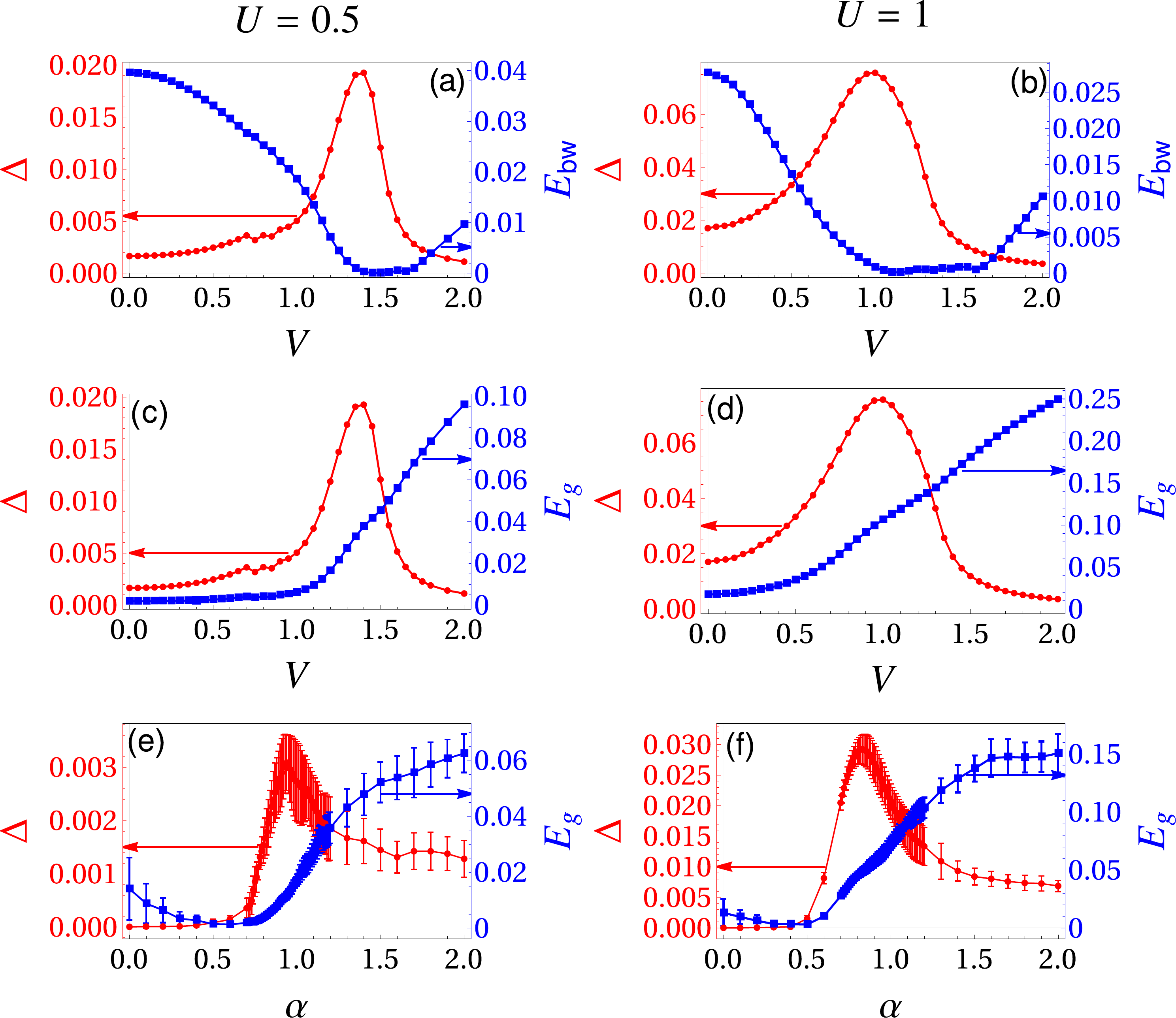}
  \caption{Superconducting order parameter $\Delta$, 
	and energy bandwidth $E_\mathsf{bw}$ of the low-lying subband for the BCS-AA model (a-b),
	and 
	single-particle energy gap $E_g$ 
	for the  
	BCS-AA (c-d) and BCS-PRBM (e-f) models with $U=0.5$ (the first column) and $U=1$ (the second column).
	(a): $\Delta$ and the bandwidth $E_\mathsf{bw}$ of the lowest lying subband in the BCS-AA model with $U=0.5$ 
	(b): Same as (a) except that $U=1$. 
	The lowest lying subband becomes almost flat at the MIT, indicating the that the (almost) diverging
	density of states also plays a role in the enhancement of $\Delta$ for the BCS-AA model.
	(c): $\Delta$ and $E_g$ for the BCS-AA model with $U=0.5$. 
	(d): Same as (c) except that $U=1$.
	The order parameter $\Delta$ peaks around the MIT, while the energy gaps increase monotonically with $V$, 
	resulting in the deviation of $E_g$ from $\Delta$.
	(e): $\Delta$ and $E_g$ in the BCS-PRBM model with $U=0.5$.
	(f): Same as (e) except that $U=1$. The error bars are from the uncertainty in the disorder averaging.
	$\Delta$ peaks around the MIT and $E_g$ increases with $\alpha$ 
	(except for small $\alpha$, where $E_g$ has large uncertainty from disorder averaging).}
	\label{fig:DeltaGaps}
\end{figure}

\begin{figure}[t]
	\centering
	\includegraphics[width=0.48\textwidth]{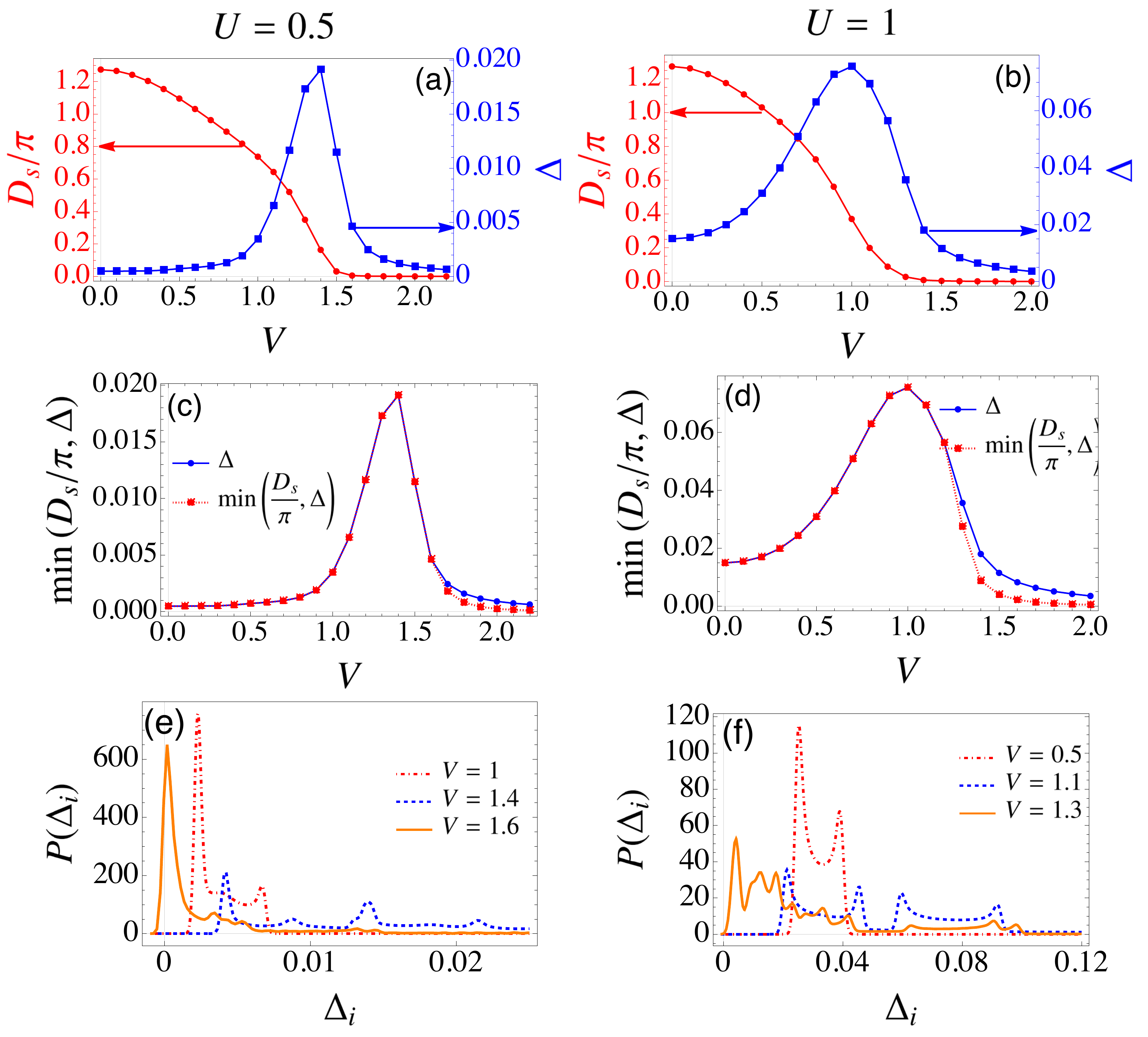}
	\caption{Superfluid stiffness $D_s/\pi$ [Eq.~\eqref{eq:Ds}], 
	order parameter $\Delta$ and distribution of the local pairing amplitude $\Delta_i$ in the BCS-AA model 
	with different interaction strengths $U=0.5$ (a,c,e) and $U=1$ (b,d,f). 
	  (a): Superfluid stiffness $D_s/\pi$ and order parameter $\Delta$ versus incommensurate potential strength $V$ in the BCS-AA model with $U=0.5$. 
	  (b): Same as (a) except $U=1$. 
	  (c): $\min\left(D_s/\pi, \Delta \right)$ along with $\Delta$ varying with $V$ for $U=0.5$. 
	  (d): Same as (c) except for $U=1$. 
	  In the delocalized phase ($V<V_c \simeq 1.4$ for $U=0.5$ and $V <V_c \simeq 1.1$ for $U=1$), $\Delta< D_s/\pi$ and the 
	enhancement of superconductivity is shown by the increasing $\Delta$ with $V$. 
	The phase fluctuations dominate in the localized phase ($V>V_c$) and $D_s/\pi<\Delta$ 
	determines the strength of the superconductivity. The multifractal enhancement of superconductivity 
	persists even incorporating the phase fluctuations. 
	  (e): Probability density of local pairing amplitude $\Delta_i$ for $U=0.5$ and different incommensurate potentials. 
	  (f): Same as (e) except for $U=1$.
	In the delocalized phase and near the MIT, $\Delta_i$ peaks around nonzero values, 
	while it peaks around $0$ in the localized phase.
	The system size is $L = 2584$ in this figure.
  }
	\label{fig:Stiffness}
\end{figure}

We find that SWQC survives at a (renormalized) single-particle MIT in the AA and PRBM models with attractive interactions.  
Our key result is that the superconducting amplitude $\Delta$ is enhanced by inhomogeneity relative to the clean case,
in a wide region around the MIT. The \emph{maximum amplitude} closely tracks the MIT for weak-to-moderate
interaction strengths, as shown in Figs.~\ref{fig:DeltaUV-AA}--\ref{fig:DeltaGaps}. 
We also compute the superfluid stiffness $D_s$ for the interacting AA model. 
We find that $D_s$ is always larger than $\Delta$, except deep in the Anderson insulator (Fig.~\ref{fig:Stiffness}).
Previous studies employing smaller system sizes also demonstrated multifractal enhancement of superconductivity in the
interacting AA model \cite{Tezuka2010, Fan2021}. In Ref.~\cite{Fan2021},
$\Delta$ and $D_s$
were computed, 
but the location of the MIT and concomitant maximization of $\Delta$ were not determined. Our calculations incorporate random Hartree
shifts \cite{Ghosal1998,Ghosal2001}, an important additional source of quantum interference (Altshuler-Aronov corrections \cite{Lee1985}). 
Our results show that when Anderson localization is prevented in low-dimensions 
(here via fine-tuned potential strengths in special models, 
but as may also occur \emph{generically} in topologically protected 2D systems \cite{Ghorashi2018,Sbierski2020,Ghorashi2020,Karcher2021}),
the rarefied nature of a wide swath of critical single-particle wave functions can strongly boost superconductivity.


\textit{Models}.---The spin-$1/2$ Aubry-Andr\'{e} model with attractive Hubbard interaction is defined via
\begin{align}
	H 
	=&\, 
	-t
	\sum_{i \sigma}
	\left(
		c^\dagger_{i \sigma}c_{i + 1 \sigma} 
		+
		c^\dagger_{i + 1 \sigma}c_{i \sigma} 
	\right)
\nonumber\\
	&\,
	+ 
	\sum_i (V_i-\mu) n_i 
	- 
	U \sum_i n_{i\uparrow}n_{i\downarrow}\,,
  \label{eq:H-AA}
\end{align}
where $c_{i \sigma}$ annihilates a spin-$\sigma \in \{\uparrow,\downarrow\}$ fermion at site $i$, 
$t$ is the nearest neighbor hopping (set to be the energy unit),  
$V_i = V \cos\left(2\pi \beta_p i \right)$ is the incommensurate potential, 
$\mu$ the chemical potential, 
$U$ the strength of attractive on-site interaction, 
and $n_i = n_{i\uparrow}+ n_{i\downarrow}$. 
We choose $\beta_p \equiv F_{p-1}/F_{p}$ to approximate the inverse golden ratio,
where $F_p$ is the $p^{\mathrm{th}}$ Fibonacci number, which is also the system size \cite{Siebesma1987}.
The system goes through a spectrum-wide MIT at $V = 2t$ without the interaction term \cite{Aubry1980, Sokoloff1985}. 
All single-particle wave functions are Anderson localized for $V>2t$,
and all of them are extended for $V < 2t$. 
All single-particle wave functions are multifractal at the critical point $V = 2t$ \cite{Evangelou1987,Siebesma1987}.
The multifractal property of the wave functions can be characterized by the scaling behavior of the inverse participation ratio (IPR)
\cite{Evers2008},
$
	P_q = \sum_i \left|\psi_i\right|^{2q} \propto L^{-\tau_q}\,,
$
with $L$ being the system size.

The dimension $\tau_q \equiv D_q (q-1)$, where 
in 1D $D_q = 1$ ($D_q = 0$) for extended (localized) states,
and $0 < D_q < 1$ for critical multifractal wave functions \cite{Evers2008}. 
Wave functions in the extended (localized) phase near the critical point can also show multifractal properties up to the scale of 
the correlation (localization) length. 
The multifractality enhancement of superconductivity can occur in a wide region close to the MIT, 
driven by critical correlations if the coherence length is shorter than the correlation or localization length \cite{Feigelman2007,Feigelman2010}.

The Hamiltonian of the spin-$1/2$ PRBM model with attractive Hubbard interactions is
\begin{equation}
	H = 
	\sum_{ij \sigma} 
	\mathsf{H}_{ij} 
	c^\dagger_{i\sigma} c_{j\sigma} 
	- 
	U \sum_i n_{i\uparrow}n_{i\downarrow} 
	- 
	\mu \sum_i n_i\,,
	\label{eq:HPRBM}
\end{equation}
where $\mathsf{H}_{ij}=\mathsf{G}_{ij}\left|i-j\right|^{-\alpha}$, with $\hat{\mathsf{G}}$ a random matrix in the orthogonal class (class AI). 
Without interactions, the system exhibits SWQC at the MIT with $\alpha=1$. 
The system is spectrum-wide extended (localized) when $\alpha<1$ ($\alpha>1$) \cite{Mirlin2000a}.


\textit{Phase Diagrams}.---In the mean-field approximation \cite{Ghosal1998,Ghosal2001,Fan2020}, the 
local superconducting order parameter $\Delta_i$ and fermion density $\left\langle n_{i}\right\rangle$ satisfy
\begin{equation}
	\Delta_{i} = -U\left\langle c_{i\downarrow}c_{i\uparrow}\right\rangle \,,
	\qquad
	\left\langle n_{i}\right\rangle =\sum_\sigma\left\langle c_{i\sigma}^{\dagger}c_{i\sigma}\right\rangle \,.
\end{equation}
We solve the systems BdG self-consistently \cite{Ghosal2001} with effective chemical potential $\tilde{\mu}_{i}=\mu+U\left\langle n_{i}\right\rangle /2$. 
The convergence condition is set so that the average difference of $\Delta_i$ and $n_i$ are smaller than $10^{-6}$ ($10^{-7}$ for small $U$) \cite{SM}. 
We focus on half-filling with $\mu=-U/2$, but the physics discussed applies to other filling factors since the whole spectrum of 
single-particle states are multifractal near the MIT.

Fig.~\ref{fig:DeltaUV-AA} shows the enhancement of the average order parameter $\Delta$ 
in the ``BCS-AA'' model [Eq.~(\ref{eq:H-AA})]. The spectrum-wide MIT persists with 
attractive interactions, and the MIT can be characterized by the second multifractal dimension 
$\tau_2$, averaged over the entire spectrum of quasiparticle states.
This shows a sharp drop from $1$ to $0$ as $V$ increases, indicating the MIT [Fig.~\ref{fig:DeltaUV-AA}(b)].
With increasing $U$, the critical incommensurate potential strength $V_\mathsf{c}$ decreases 
and the Anderon insulator phase is enlarged, Fig.~\ref{fig:DeltaUV-AA}(a). 
$\Delta$ is enhanced by the multifractal wave functions near the transition, and the maximal $\Delta$ for fixed $U$ 
follows the MIT curve $V_c(U)$ for weak and moderate interactions.
When the incommensurate potential strength $V$ is weak, the order parameter is determined by BCS theory with $\Delta \sim \exp{(-1/U\nu)}$, 
with $\nu$ the density of states at the Fermi point.
As $V$ increases, $\Delta$ increases significantly and peaks around the MIT, e.g.\ 
$\Delta(V_\mathsf{c})$ is more than $10$ times larger than $\Delta\left(V=0 \right)$ for $U=0.5$.
The order parameter amplitude decreases in the Anderson insulator phase due to the combination of localization and Altshuler-Aronov effects \cite{Ma1985,Lee1985}. 
The enhancement ratio $\Delta(V_\mathsf{max})/\Delta(V=0)$ decreases as $U$ increases and the strongest enhancement curve
deviates from the MIT curve at strong interaction.

Apart from inducing SWQC of the wave functions at $V_\mathsf{c}$, the potential in the BCS-AA model 
additionally generates the interaction-dressed Hofstadter energy spectrum. 
Band flattening near half-filling plays a role in the enhancement of the order parameter seen here, 
and the maximum $\Delta$ also occurs close to the band flattening point, Figs.~\ref{fig:DeltaGaps}(a),(b). 
The density of states is much larger at the band flattening regions, but 
the average order parameter deviates significantly from the homogeneous BCS prediction 
$\Delta \sim \exp{(-1/U\nu)}$, except for small $V$ ($V<0.5$ for $U=0.5$). 
Multifractal enhancement \emph{without} band flattening is observed in the BCS-PRBM model (described below).

The single-particle wave functions become more and more rarefied with increasing $V$, resulting
in a stronger binding energy between paired electrons occupying the same spatial orbital, and thus increasing the 
spectral gap $E_g$. In the strong-localization limit, the pairing energy is given by $U P_2(E)$, 
with $P_2(E)$ the IPR of the localized state. The energy gap of the BCS-AA model is then given by half the 
pairing energy $E_g =P_2(E_0)U/2$ \cite{Ghosal2001,SM},
with $E_0$ the energy of the lowest quasiparticle state.
Unlike $\Delta$, the gap $E_g$ in our numerics always increases with $V$, 
and is much larger than $\Delta$ for finite $V$ [Fig.~\ref{fig:DeltaGaps}(c),(d)].
Thus while the pairing energy of more localized states is larger than extended ones,
the average amplitude $\Delta$ is suppressed in the insulator by the strong fluctuations
of $\Delta_i$ in space and the loss of multifractal enhancement. 
The increasing of $E_g$ into the localization regime is consistent with previous studies 
indicating that the energy gap increases with the inverse of localization length \cite{Ghosal2001,Bouadim2011}.

Fig.~\ref{fig:Delta-PRBM} demonstrates the enhancement of $\Delta$ in the ``BCS-PRBM'' model [Eq.~(\ref{eq:HPRBM})]. 
Fig.~\ref{fig:Delta-PRBM}(a) is a contour plot of $\Delta$ as a function of $U$ and the hopping exponent $\alpha$,
near the interaction-dressed MIT. The order parameter $\Delta$ takes its largest value close to the MIT curve obtained by fitting 
$\tau_2$ of the lowest energy quasiparticle state. The change of $\tau_2$ from the extended phase ($\tau_2\sim1$) to localized phase 
($\tau_2\sim0$) with $\alpha$ is much slower in the BCS-PRBM model, compared to that in BCS-AA model, resulting in a much broader critical region.
The SWQC wave functions survive in the presence of attractive interactions and pairing, but the $\tau_2$ of the quasiparticle states are affected 
differently for different states. The lowest-lying quasiparticle states are the best indicator for the MIT and $\Delta$ enhancement,
as these are most involved in pairing. The enhancement always occurs in the critical region, indicated by the drop of $\tau_2$ in Figs.~\ref{fig:Delta-PRBM}(b) and (c). 
The spectral gap $E_g$ in the BCS-PRBM model shows similar behavior as that in the BCS-AA, increasing with $\alpha$ to the localized phase 
[Fig. \ref{fig:DeltaGaps}(e),(f)]. 
In the localized phase, $E_g$ is also approximately proportional to $P_2(E_0)$ \cite{SM}.
Different the from BCS-AA model, there is no significant change of the density of states across the MIT 
in the BCS-PRBM model, and the critical wave functions are the only factor responsible for enhancing $\Delta$.

\textit{Superfluid Stiffness}.---Strong phase fluctuations 
in low dimensions can demolish superconductivity even if the pairing amplitude remains finite. 
In a spatially inhomogeneous system, regions with small $\Delta_i$ enhance phase fluctuations.
The phase rigidity of a superconductor can be described by the superfluid stiffness 
\cite{Scalapino1993, Trivedi1996}. 
In a gapped one-dimensional system, the superfluid stiffness is determined by
\begin{equation}
	\frac{D_s}{\pi} 
	= 
	\Pi_{xx}^R\left(q_x=0,\omega\to 0\right)-\left\langle K_{x}\right\rangle \,.
	\label{eq:Ds}
\end{equation}
Here $\Pi_{xx}^R$ is the retarded current-current correlation function and $K_x$ is the kinetic energy density. 
The above $q_x=0$ and $\omega \to 0$ limit gives the 
Drude weight $D$; 
it can be shown that $D_s=D$ at zero temperature for gapped systems \cite{Scalapino1993, Giamarchi1995}. 
We employ Eq.~\eqref{eq:Ds} to evaluate 
$D_s$ in 
the BCS-AA model 
with s-wave pairing.

Fig.~\ref{fig:Stiffness} shows the superfluid stiffness $D_s/\pi$ and order parameter $\Delta$ in BCS-AA model. 
The superfluid stiffness $D_s$ decreases monotonically with increasing incommensurate potential, 
while $\Delta$ peaks around MIT, Figs.~\ref{fig:Stiffness}(a,b).
The minimum of $D_s/\pi$ and $\Delta$ determines the strength of the superconductivity.  
We plot $\min{\left( D_s/\pi, \Delta \right)}$ in Figs.~\ref{fig:Stiffness}(c,d). 
In the delocalized phase, $\Delta$ is much smaller than $D_s/\pi$ and becomes comparable with $D_s/\pi$ near the MIT. 
Only in the localized phase, $D_s/\pi$ becomes smaller than $\Delta$. 
The distribution of the local pairing amplitude $\Delta_i$ is illustrated in Figs.~\ref{fig:Stiffness}(e,f). 
The probability density of $\Delta_i$ peaks at nonzero values in the delocalized phase and near the MIT; by contrast, 
it peaks around $0$ in the localized phase. This indicates that the finite average $\Delta$ in the localized phase 
is due to rare regions with large values of $\Delta_i$.


\textit{Conclusion}.---
We have shown
that the pairing amplitude for superconductivity is enhanced by SWQC
in the BCS-AA and -PRBM models.
The maximal enhancement tracks the MIT
in both models. 
The enhancement survives phase fluctuations at zero temperature, 
supported by the superfluid stiffness data for the BCS-AA model.
Although true superconductivity does not occur in 1D \cite{Giamarchi},
SWQC also emerges in 2D systems \cite{Karcher2021}.
Strong spatial fluctuations observed in $\Delta(\vex{r})$ in the high-$T_c$ cuprate superconductors \cite{DavisReview} 
may realize SWQC for nodal quasiparticles \cite{Ghorashi2020}. 

Generalized AA models 
have been proposed \cite{Soukoulis1982, DS1988, Biddle2010, Ganeshan2015, Li2017, Yao2019, Yao2020} and studied in recent experiments \cite{Luschen2018, Kohlert2019, An2021}. 
The pairing amplitude enhancement could also be examined in these systems when the Fermi level is tuned close to the mobility edge.

We thank Ferdinand Evers and Nandini Trivedi for very useful discussions.
This work was supported by the Welch Foundation Grant No.~C-1809.

\end{document}


\title{Supplemental Material: Enhanced Amplitude for Superconductivity
due to Spectrum-wide Wave Function Criticality 
in Quasiperiodic and Power-law Random Hopping Models}
\author{Xinghai Zhang}
\affiliation{Department of Physics and Astronomy, Rice University, Houston, Texas
77005, USA}
\author{Matthew S.\ Foster}
\affiliation{Department of Physics and Astronomy, Rice University, Houston, Texas
77005, USA}
\affiliation{Rice Center for Quantum Materials, Rice University, Houston, Texas
77005, USA}
\maketitle
\tableofcontents

\section{Convergence Condition}

In our numerics, we solve the Bogoliubov-de Gennes equations self-consistently with the inhomogeneous pairing and Hartree shift,
\begin{equation}
	\Delta_i 
	= 
	-
	U \left<c_{i\downarrow} c_{i\uparrow}\right>\,,\qquad \left<n_i\right> 
	= 
	\sum_i \left<c^\dagger_{i\sigma} c_{i\sigma}\right>\,.
	\label{eq:Deltai}
\end{equation}
The error of each iteration is defined as
\begin{eqnarray}
  \Delta_\mathsf{err} &=& \frac{1}{L} \sqrt{\sum_{i=1}^L \left(\Delta_i^{m+1}-\Delta_i^{m}\right)^2}\,,\\
  n_\mathsf{err} &=& \frac{1}{L} \sqrt{\sum_{i=1}^L \left(n_i^{m+1}-n_i^{m}\right)^2}\,.
  \label{eq:Err}
\end{eqnarray}
Here $\Delta_i^{m}$ and $n_i^{m}$ are the order parameter and Hartree shift at $i$-th site in the $m$-th iteration, and $L$ is the number of sites. 
We terminate the iteration when $\max{\left( \Delta_\mathsf{err}, n_\mathsf{err} \right)}$ is smaller than $10^{-6}$, or $10^{-7}$ (when $U$ and $\Delta$ are small). 
In the extended phase, the numerics usually converges in around $30$ or $50$ iterations. 
In the localized phase, the maximum iteration is set to be $100$ to reduce computational consumption and 
$\max{\left\{ \Delta_\mathsf{err}, n_\mathsf{err} \right\}}$ is around $2\sim 3 \times 10^{-6}$ when terminated.

\section{Multifractal Properties of the BCS-Aubry-Andr\'{e} Model}

\begin{figure}[th]
	\centering
	\includegraphics[width=0.98\textwidth]{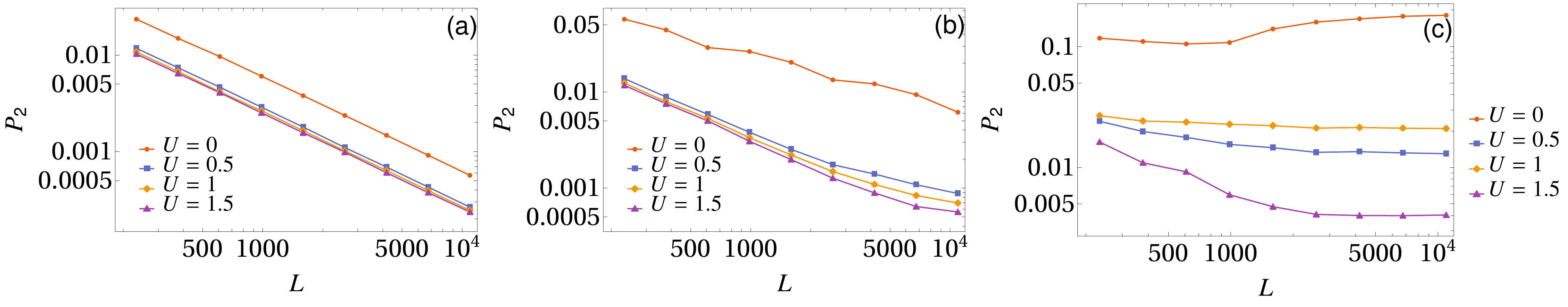}
	\caption{Inverse participation ratio $P_2$ around the Anderson-localization transition in the BCS-AA model. 
	(a): In the extended phase with $V\approx V_{c}-0.1$. 
	(b): Close to transition point $V_c$. 
	(c): In the localized phase.}
	\label{fig:P2-AA}
\end{figure}

In Fig. \ref{fig:P2-AA}, we show the average inverse participation ratio (IPR) of the BCS-AA model,
\begin{equation}
	P_2 
	=
\frac{1}{L} \sum_{n=1}^{L} \sum_{i = 1}^L \left|\psi^n_i\right|^4.
	\label{eq:P2}
\end{equation}
Here $\psi^n$ is the $n$-th quasiparticle state with positive energy for the BCS-AA Hamiltonian. 
We can see that $P_2\sim L^{-1}$ in the extended phase, 
$P_2\sim L^0$ in the localized phase, 
and 
$P_2\sim L^{-\tau_2}$ with $0<\tau_2<1$ close to the critical point.

\begin{figure}[th]
  \centering
  \includegraphics[width=0.58\textwidth]{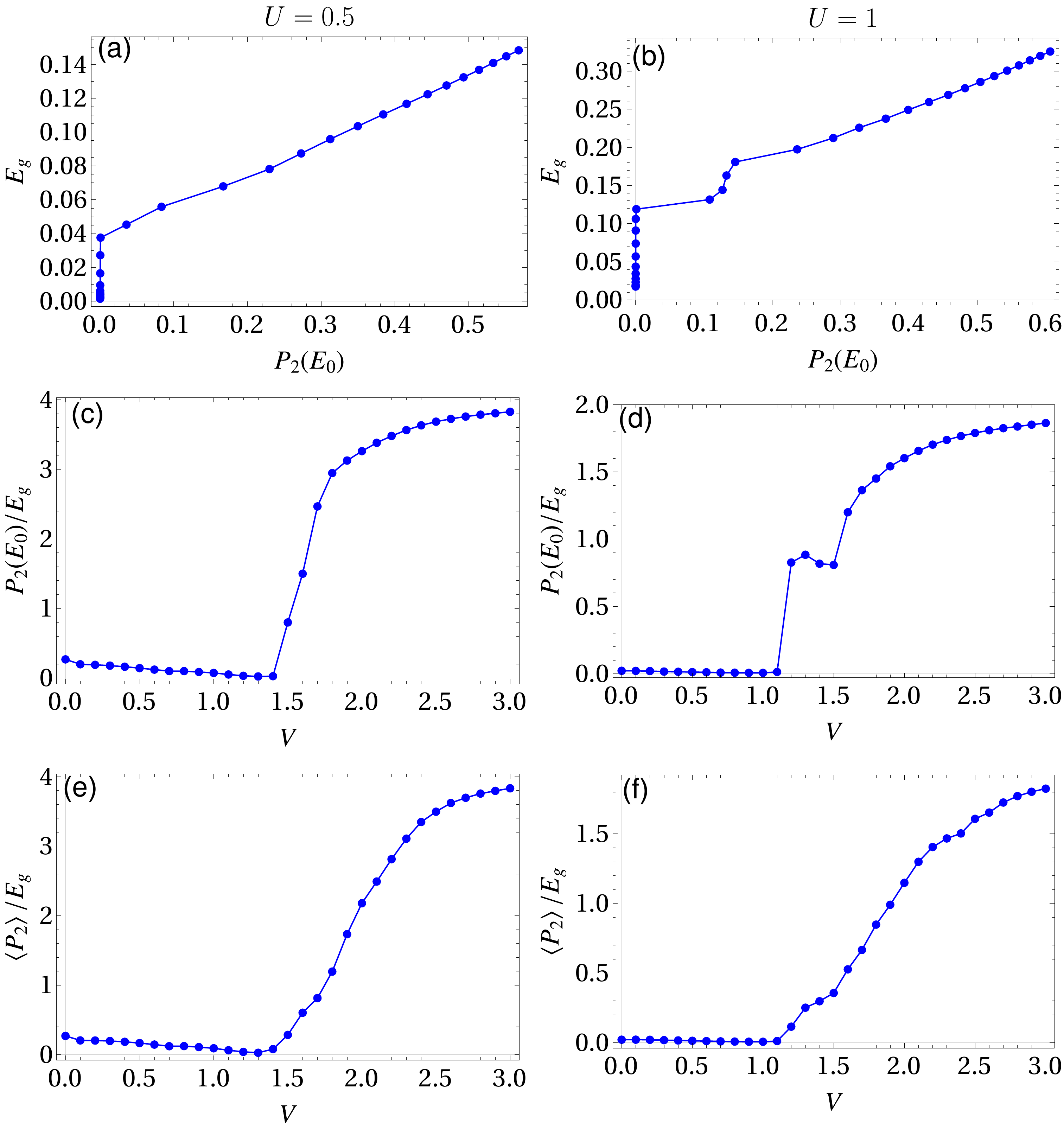}
  \caption{Single-particle energy gap $E_g$ and inverse participation ratio $P_2(E_0)$ of the lowest energy quasiparticle state in the BCS-AA model. 
	(a): $E_g$-$P_2(E_0)$ with $U=0.5$. $E_g$ is proportional to $P_2(E_0)$ in the localized phase and $E_g\approx P_2(E_0)U/2$ in the strong-localization limit. 
	(b): $E_g$-$P_2(E_0)$ with $U=1$. 
	(c): $P_2(E_0)/E_g$ as a function of $V$ with $U=0.5$. This approaches $2/U=4$ in the strong-localization limit. 
	(d): $P_2(E_0)/E_g$ as a function of $V$ with $U=1$. 
	(e): $\left<P_2\right>/E_g$-$V$ at $U=0.5$, where $\left<\cdots\right>$ denotes the spectrum-wide average. 
	(f): the same as (e) except for $U=1$.}
  \label{fig:Egap-P2-AA}
\end{figure}

The single-particle energy gap $E_g$ is half of the pairing energy of a singlet pair of two electrons occupying the spatial orbital eigenstate 
with the lowest energy. In the strong localization limit, the spatial overlap [given by the Chalker correlation function, Eq.~(5) in the main text] 
of two different single-particle orbital states with nearby energies vanishes, since these are typically exponentially localized in a small, distant regions. 
The paired electrons occupy the same strongly localized orbital state, which is nonzero only at a small portion of the sites.
The energy gap $E_g$, i.e.\ half of the pairing energy is then given by \cite{SM-Ghosal2001},
\begin{equation}
	E_g = P_2(E_0)U/2.
	\label{eq:Eg-P2}
\end{equation}
The inverse of the IPR $1/P_2$ gives a good approximation for the localization length in the localized phase.
Fig.~\ref{fig:Egap-P2-AA} shows the relation between the single-particle energy gap $E_g$ and the inverse participation ratio $P_2(E_0)$ 
of the lowest-energy quasiparticle state. We can see that $P_2(E_0)/E_g$ approaches $2/U$ in the strong-localization limit $V \gg V_c$.

\begin{figure}[th]
  \centering
  \includegraphics[width=0.98\textwidth]{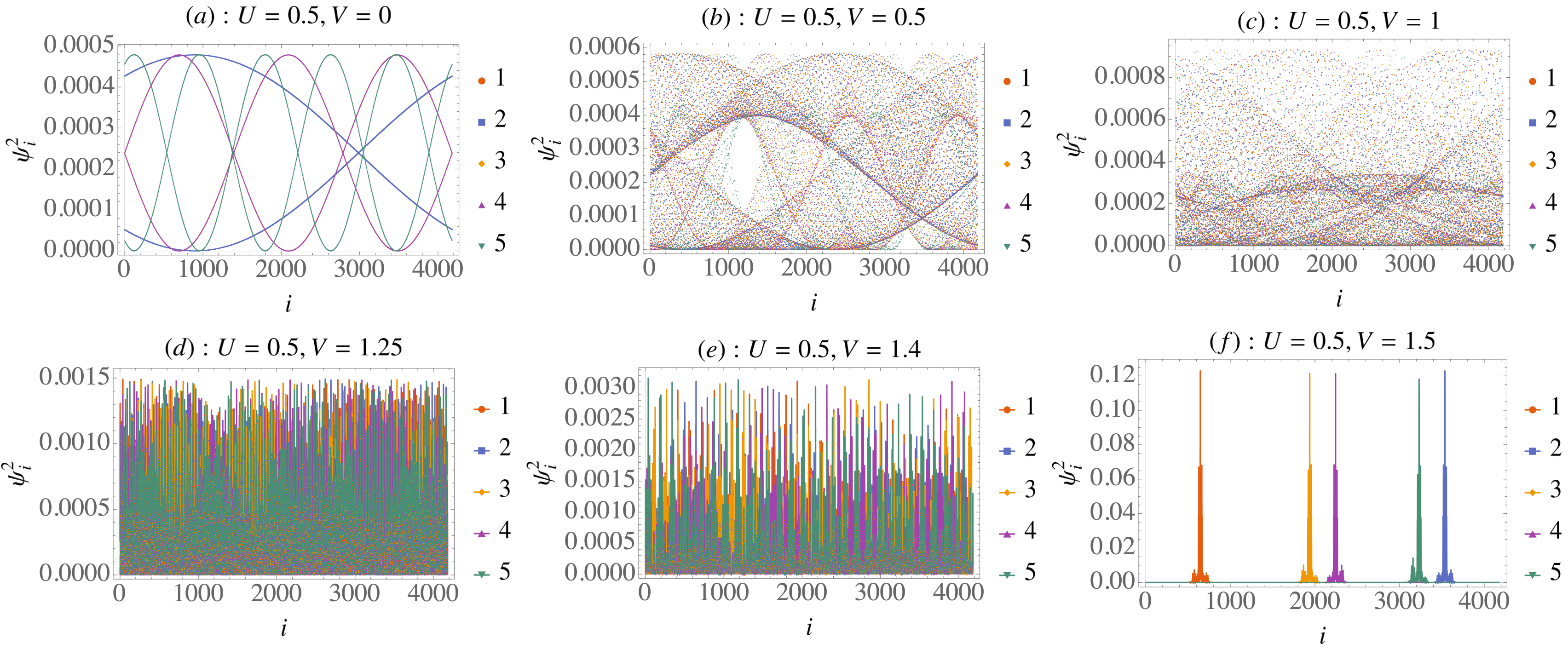}
  \caption{Probability density $|\psi_i|^2$ of five low-lying quasiparticle eigenstates in the BCS-AA model with $U=0.5$. 
	The number labels indicate the quasiparticle states with the $i$-th lowest energy. 
  The Anderson-localization transition occurs around $V\approx 1.4$. 
	The wave functions, especially the ones with small $V$, show a rapid oscillation with period of $2$ lattice sites and the probability densities appear 
	double-valued. This is due to the $k_F$-oscillation since the system is doped at half-filling with Fermi momentum $k_F=\pm \frac{\pi}{2 a}$. }
  \label{fig:Psi2-AA-U=0.5}
\end{figure}

\begin{figure}[th]
  \centering
  \includegraphics[width=0.98\textwidth]{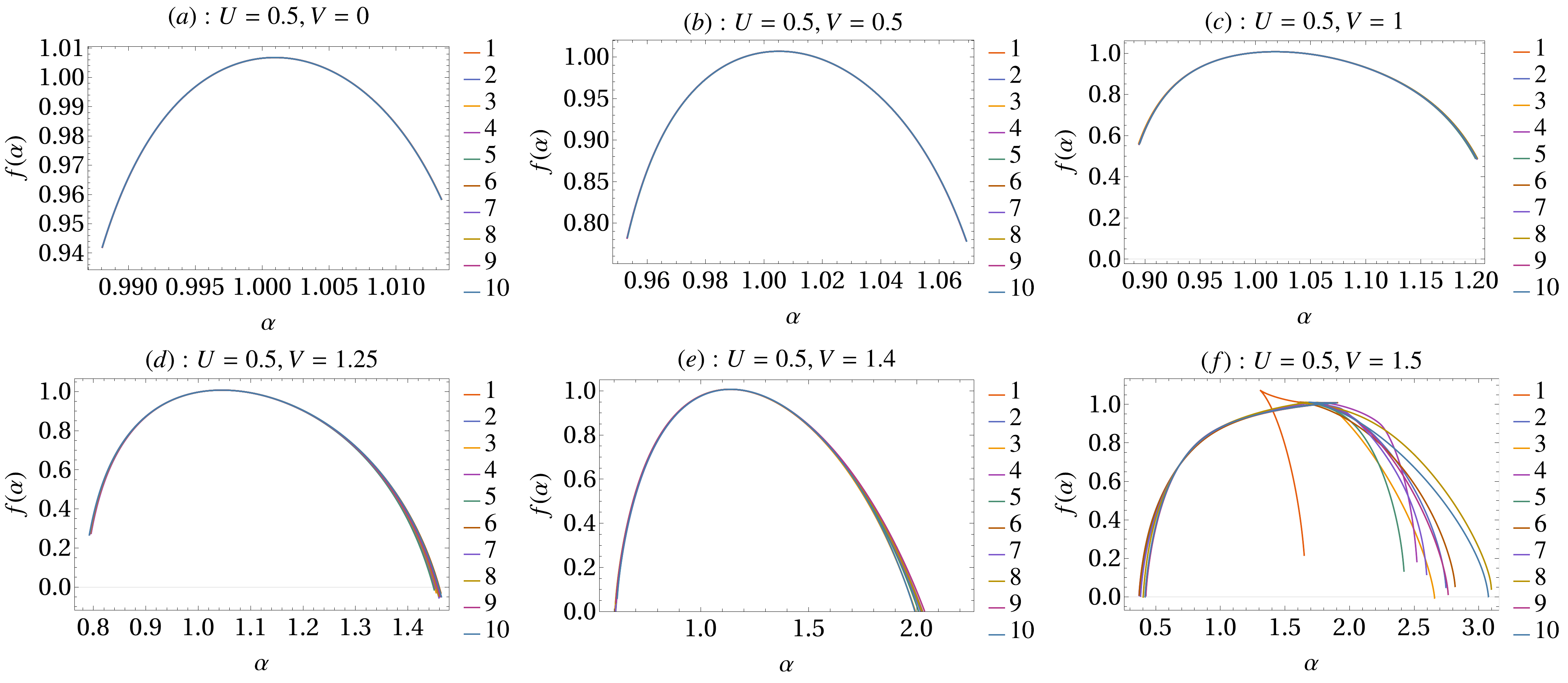}
  \caption{The singularity spectrum $f(\alpha)$ of $10$ low-lying quasiparticle states in the BCS-AA model with $U=0.5$. 
	For localized states, the $q<0$ data are not reliable due to the singularity caused by $\psi_i\approx 0$ for points away 
	from the localization center, and this is reflected in the discontinuity in $f(\alpha)$ curve with $\alpha>\alpha_\mathsf{max}$.}
	\label{fig:alphaf-U=0.5}
\end{figure}

\begin{figure}[th]
  \centering
  \includegraphics[width=0.98\textwidth]{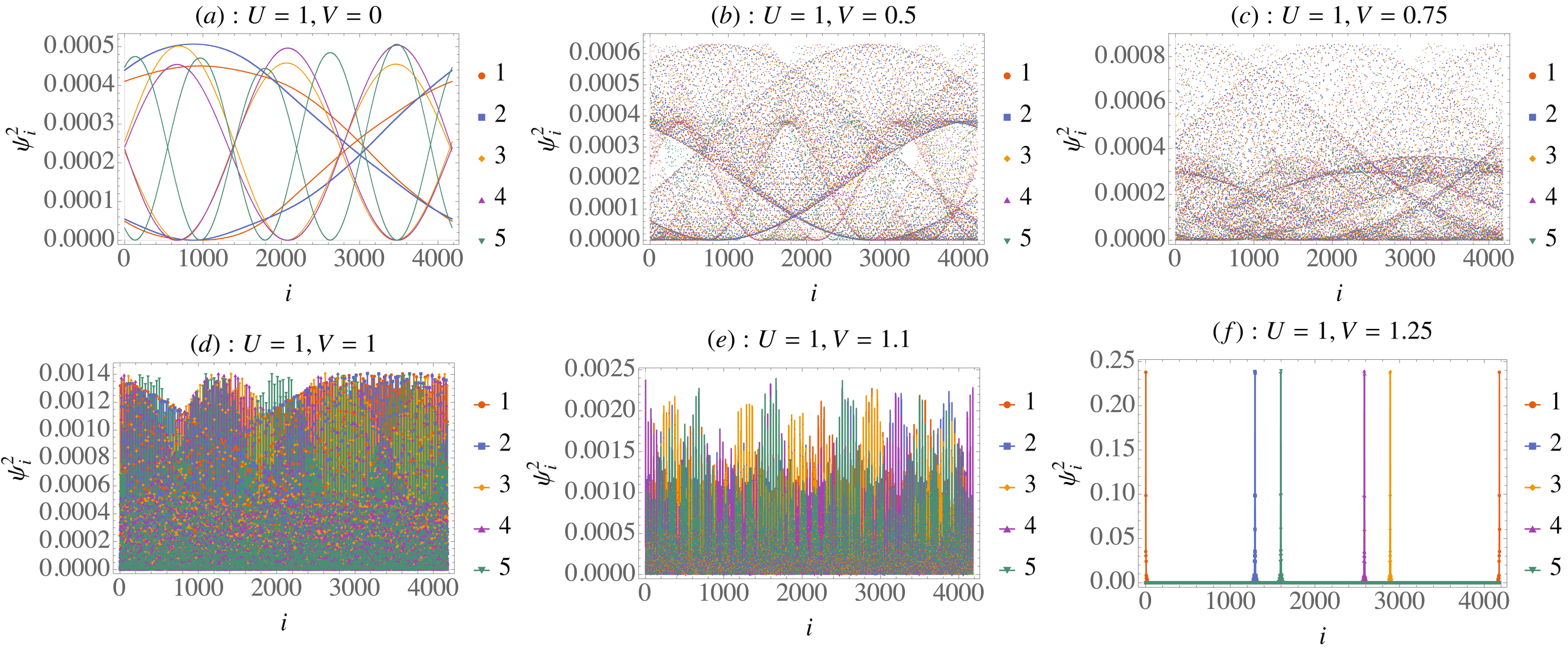}
  \caption{Probability density $|\psi_i|^2$ of low-lying quasiparticle states in the BCS-AA model with $U=1$. 
  The Anderson-localization transition occurs around $V\approx 1.1$. 
  The wave functions at small $V$ also show $k_F$-oscillation, as explained in Figure \ref{fig:alphaf-U=0.5}.}
  \label{fig:Psi2-AA-U=1}
\end{figure}

In Figs.~\ref{fig:Psi2-AA-U=0.5} and \ref{fig:Psi2-AA-U=1}, we plot the spatial probability density $|\psi_i|^2$ of $5$ low-lying
quasiparticle eigenstates of the BCS-AA model with $U=0.5$ and $U=1$, for different incommensurate potential strengths $V$
through the Anderson-localization transition. Without the incommensurate potential, all the single
particle wave functions are plane waves [Fig.~\ref{fig:Psi2-AA-U=0.5}(a) and \ref{fig:Psi2-AA-U=1} (a)]. 
As the potential strength $V$ increases, the wave functions are no longer pure plane waves and they begin to show some 
multifractal features at short distances [Figs.~\ref{fig:Psi2-AA-U=0.5}(b-d) and \ref{fig:Psi2-AA-U=1} (b-d)]. 
Near the critical point, the wave functions are multifractal in the whole space, and the probability densities of eigenstates 
with similar energies overlap strongly in space [Fig.~\ref{fig:Psi2-AA-U=0.5}(e) and \ref{fig:Psi2-AA-U=1} (e)]. 
In the localized phase, the wave functions are localized in a narrow region of space [Fig.~\ref{fig:Psi2-AA-U=0.5}(f) and \ref{fig:Psi2-AA-U=1}(f)].

\begin{figure}[th]
  \centering
  \includegraphics[width=0.98\textwidth]{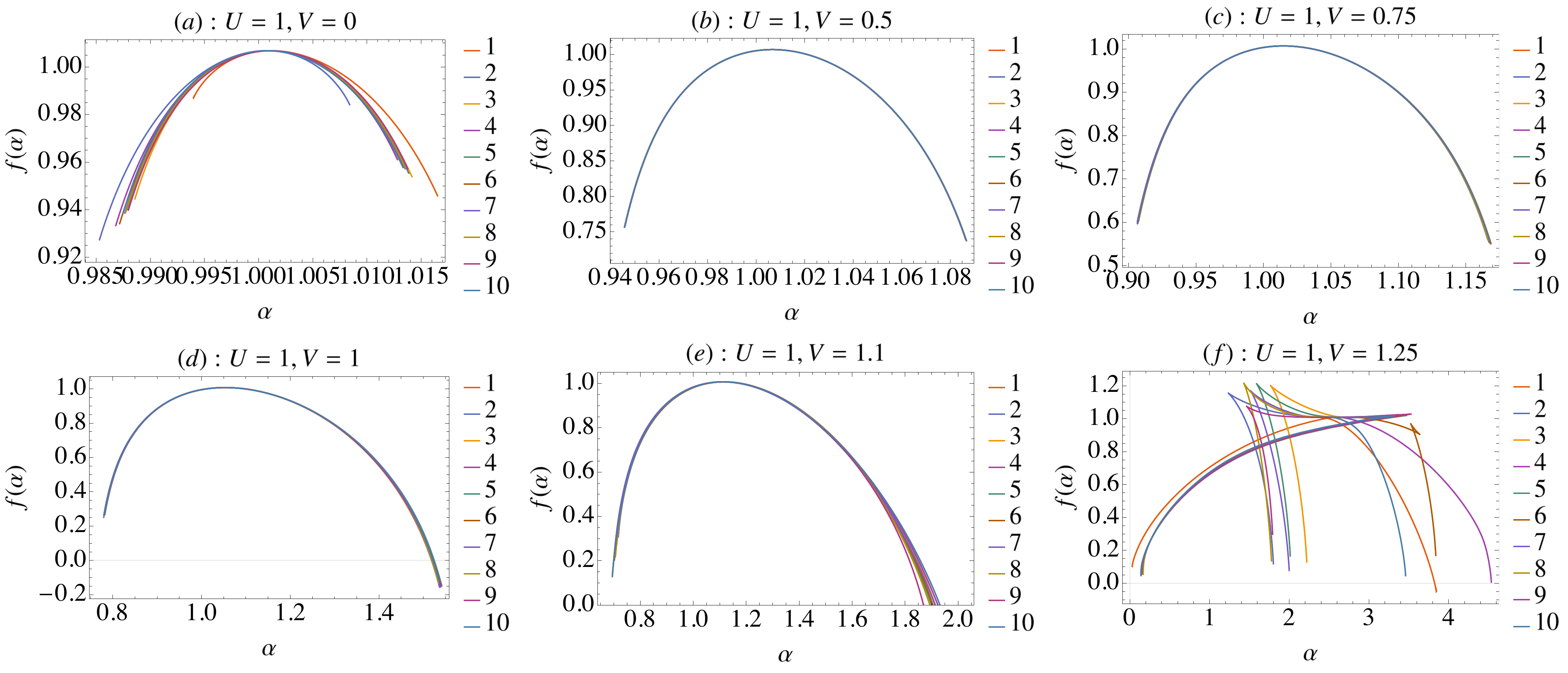}
  \caption{The singularity spectrum $f(\alpha)$ of the $10$ low-lying quasiparticle states in BCS-AA model with $U=1$. 
	For localized states, the $q<0$ data are not reliable due to the singularity caused by $\psi_i\approx 0$ points away from the localization center,
	and this is reflected in the discontinuity in $f(\alpha)$ curve with $\alpha>\alpha_\mathsf{max}$.}
  \label{fig:alphaf-U=1}
\end{figure}

In the multifractal analysis of a system, the multifractal dimension $\tau_q$ is the defined as the exponent with $P_q\sim L^{-\tau_q}$ in the 
limit that the system size $L$ goes to $\infty$. The multifractal analysis can also be performed for a single wave function. The inverse participation 
ratio with the system divided into $L/b$ boxes of size $b$ is defined as \cite{SM-Halsey1986, SM-Siebesma1987},
\begin{equation}
	P_q(b) 
	= 
	\sum_{n=1}^{L/b}
	\left( \sum_{i=1}^b \left|\psi_{(n-1)b+i}\right|^2 \right)^q 
	\propto 
	b^{\tau_{q}}.
  \label{eq:Pqb}
\end{equation}
The exponent $\tau_q$ can be expressed as
\begin{equation}
	\tau_q = q\alpha-f(\alpha)\,,\qquad \alpha = \frac{d\tau_q}{dq}\,.
  \label{eq:tauq}
\end{equation}
The singularity spectrum $f(\alpha)$ is the fractal dimension of points in a wave function that scales like 
$\left|\psi_i\right|^2\sim L^{-\alpha}$ \cite{SM-Halsey1986, SM-Evers2008}. 
In Figs.~\ref{fig:alphaf-U=0.5} and \ref{fig:alphaf-U=1}, we present the singularity spectrum of $10$ low-lying 
quasiparticle states in the BCS-AA model with $U=0.5$ and $U=1$, at different incommensurate potential strengths $V$ through the Anderson-localization transition. 
For plane waves, $f(\alpha) = d \, \delta_{\alpha,d}$ in $d$ dimensions, and we can see that $f(\alpha)$ takes nonzero value around $1$ at $\alpha\approx1$ for wave functions 
in the BCS-AA model without the incommensurate potential [Figs.~\ref{fig:alphaf-U=0.5}(a) and \ref{fig:alphaf-U=1}(a)]. 
Close to the critical point, the wave functions are multifractal and $f(\alpha)$ shows approximately parabolic structure in a large range of $\alpha$. 
In the localized phase, $f(\alpha)$ is only defined up to the localization length in the $f'(\alpha)>0$ side. $f'(\alpha)$ values are obtained from 
$\tau_q$ with $q>0$, and $\tau_q$ is not well-defined for $q<0$ in the localized regime since $\psi_i \approx 0$ away from the localization center.

\section{Multifractal Properties of the BCS-PRBM Model}

\begin{figure}[ht]
  \centering
  \includegraphics[width=0.88\textwidth]{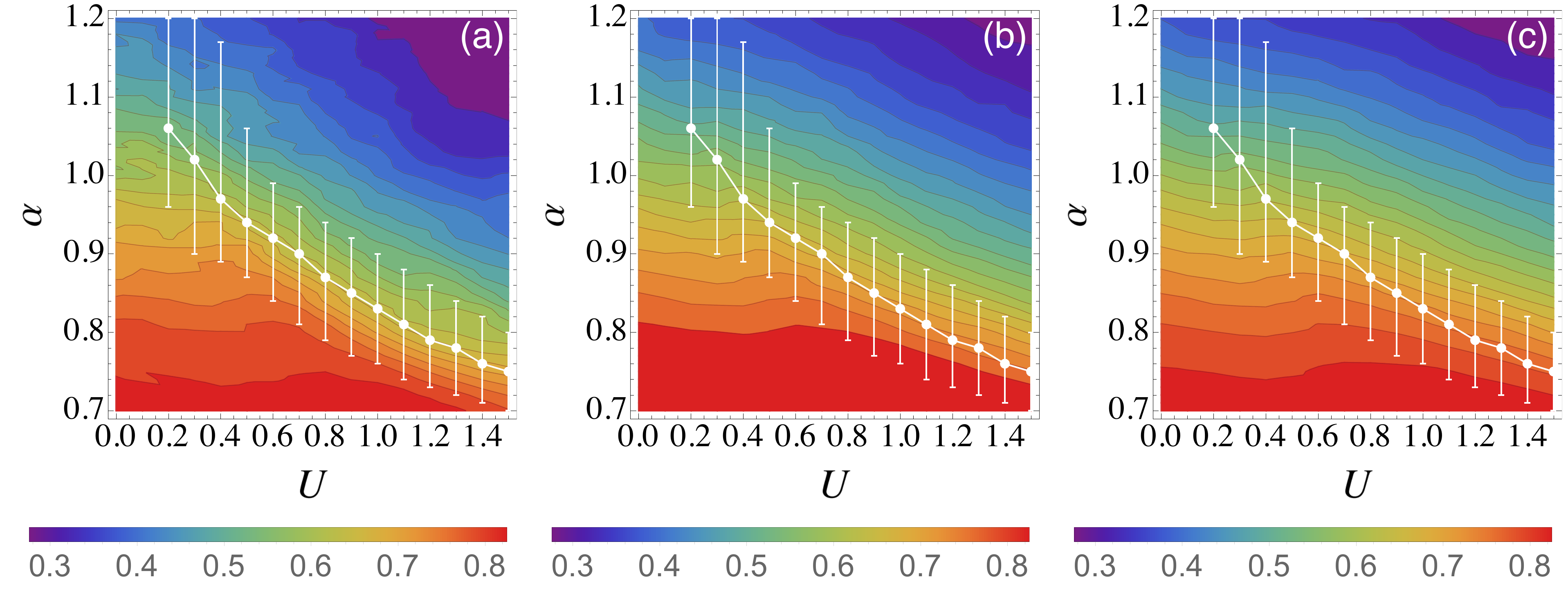}
  \caption{$\left<\tau_2\right>$ of low-lying quasiparticle states in BCS-PRBM model, where $\left<\cdots\right>$ represents the disorder averaging of $20$ samples of size $L=2000$. 
	(a): $\left<\tau_2\right>$ of the lowest energy quasiparticle states. 
	(b): $\left<\tau_2\right>$ of the $10$ low-lying quasiparticle states. 
	(c): $\left<\tau_2\right>$ of the $20$ low-lying quasiparticle states.}
  \label{fig:tau2-PRBM}
\end{figure}

\begin{figure}[ht]
  \centering
  \includegraphics[width=0.58\textwidth]{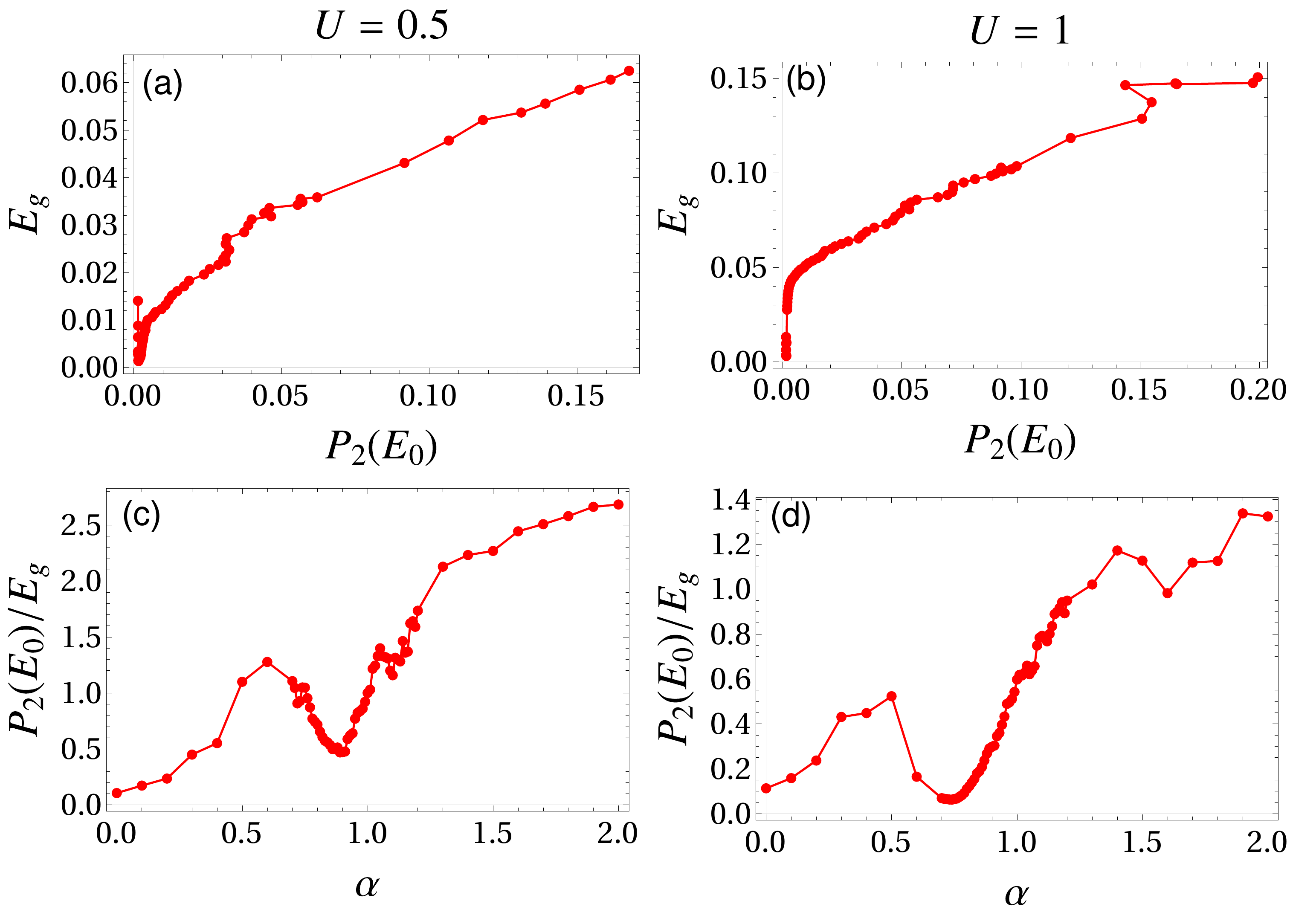}
  \caption{Single-particle energy gap $E_g$ and inverse participation ratio $P_2(E_0)$ of the lowest-energy quasiparticle state in the BCS-PRBM model. 
	(a): $E_g$-$P_2(E_0)$ with $U=0.5$. $E_g$ is proportional to $P_2(E_0)$ in the localized phase. 
	(b): $E_g$-$P_2(E_0)$ with $U=1$. 
	(c): $P_2(E_0)/E_g$ as a function of $\alpha$ with $U=0.5$. This does not reach $2/U=4$ even at $\alpha = 2$, since the localization length $1/P_2(E_0)\gg 1$ there.
	(d): $P_2(E_0)/E_g$ as a function of $\alpha$ with $U=1$.}
  \label{fig:Egap-P2-PRBM}
\end{figure}

In Fig.~\ref{fig:tau2-PRBM}, we show $\left<\tau_2\right>$ for the disorder-averaged lowest energy state, 
$0.5\%$, and $1\%$ average of the low-lying quasiparticle states for BCS-PRBM model. 
The most enhanced superconductivity curve tracks the change of $\tau_2$ (except for small $U$, where $\Delta$ is not very accurate).

Fig.~\ref{fig:Egap-P2-PRBM} shows the relation between $E_g$ and $P_2(E_0)$. $P_2(E_0)/E_g$ is not close to $2/U$ in the localized phase since the 
strong-localization limit is still not reached at $\alpha=2$. However, the ratio at $U=1$ is approximately half of that at $U=0.5$, implying that the relation 
$E_g\sim P_2(E_0)U/2$ gives a good qualitative estimation of the energy gap $E_g$ in BCS-PRBM model.

\begin{figure}[ht]
  \centering
  \includegraphics[width=0.98\textwidth]{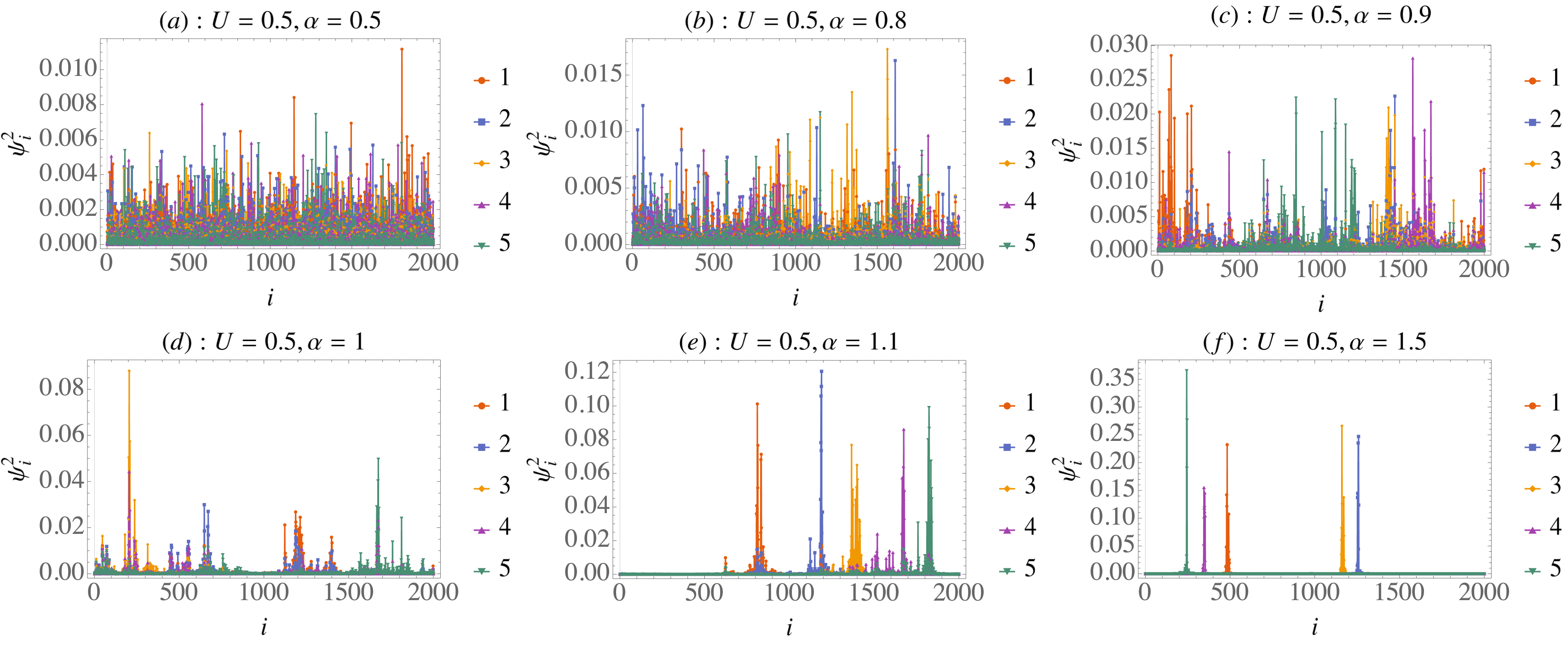}
  \caption{Probability density $|\psi_i|^2$ of $5$ low-lying quasiparticle states around the Anderson-localization transition in the BCS-PRBM model with $U=0.5$. 
	The Anderson-localization transition occurs around $\alpha\approx 0.96$.}
  \label{fig:Psi2-PRBM-U=0.5}
\end{figure}

\begin{figure}[ht]
  \centering
  \includegraphics[width=0.98\textwidth]{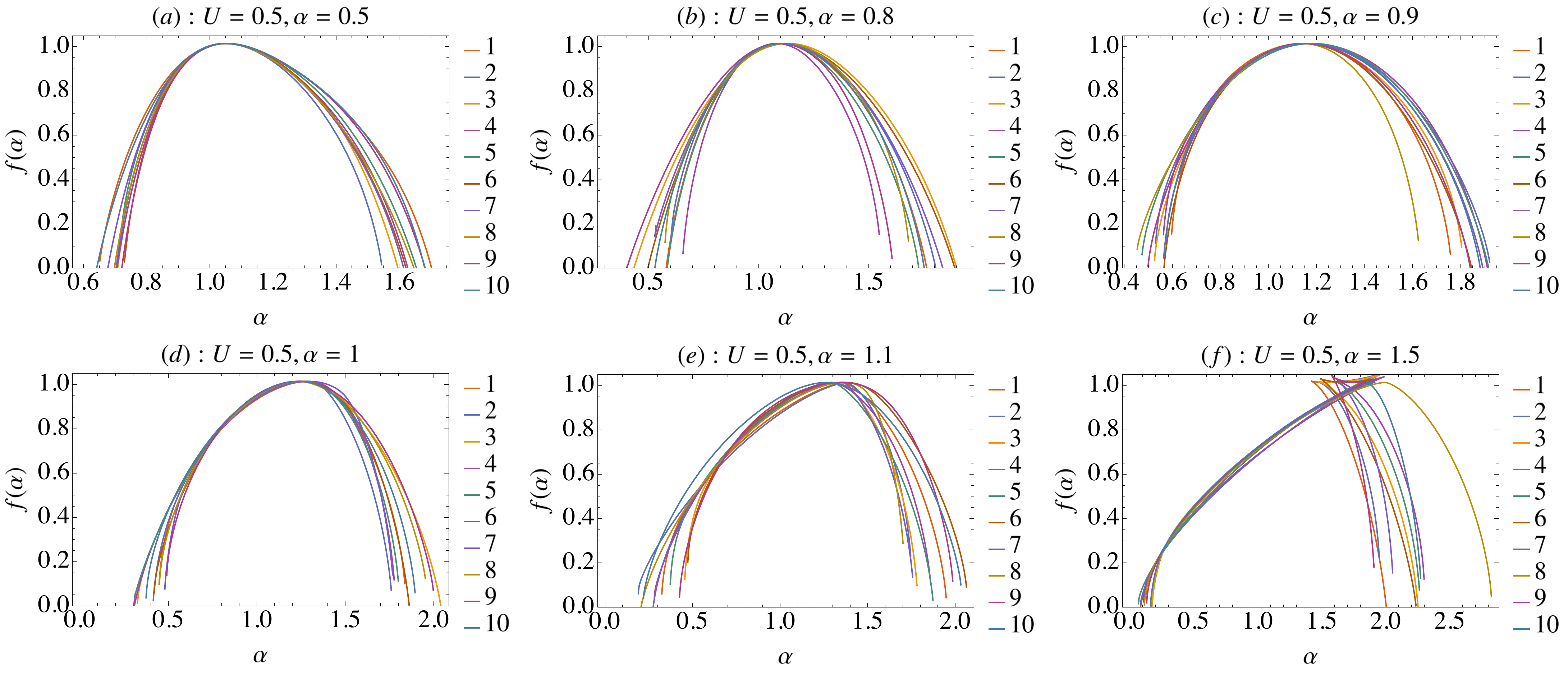}
  \caption{Singularity spectrum of the BCS-PRBM model with $U=0.5$, with $\alpha$ tuned through the Anderson-localization transition.}
  \label{fig:alphaf-PRBM-U=0.5}
\end{figure}

\begin{figure}[ht]
  \centering
  \includegraphics[width=0.98\textwidth]{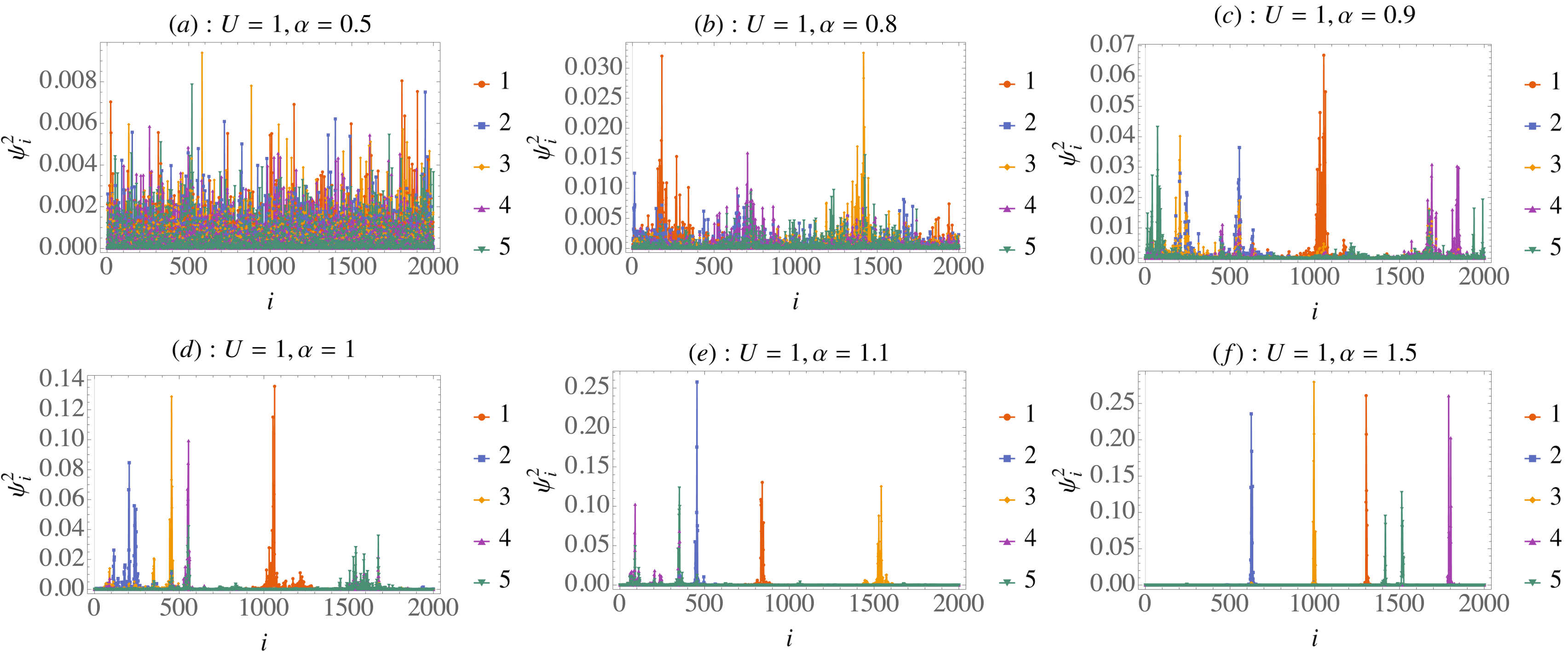}
  \caption{Probability density $|\psi_i|^2$ of $5$ low-lying quasiparticle states in the BCS-PRBM model with $U=1$. 
	The Anderson-localization transition occurs at around $\alpha\approx 0.86$.}
  \label{fig:Psi2-PRBM-U=1}
\end{figure}

\begin{figure}[ht]
  \centering
  \includegraphics[width=0.98\textwidth]{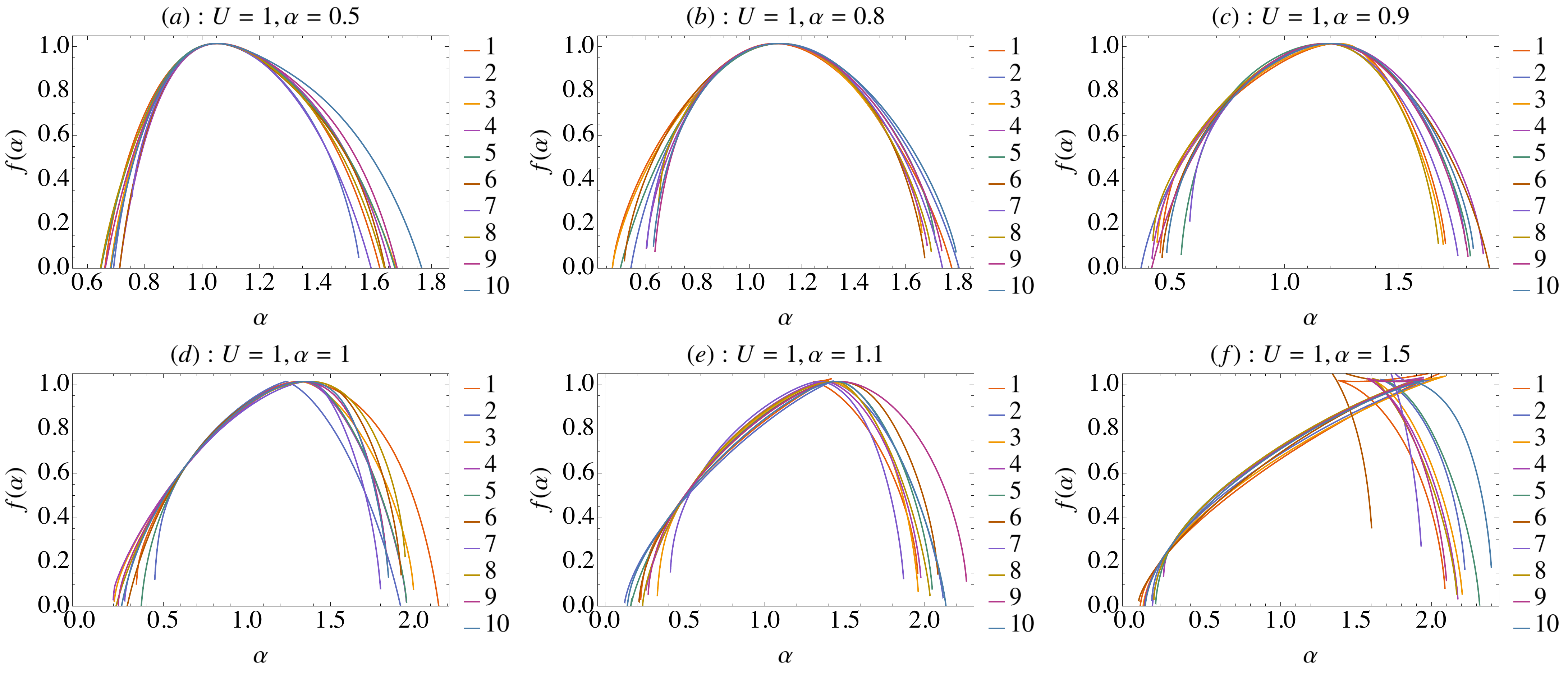}
  \caption{Singularity spectrum of the BCS-PRBM model with $U=1$ through the Anderson-localization transition.}
  \label{fig:alphaf-PRBM-U=1}
\end{figure}

In Figs.~\ref{fig:Psi2-PRBM-U=0.5} and \ref{fig:Psi2-PRBM-U=1}, we show $|\psi_i|^2$ for the BCS-PRBM model with $U=0.5$ and $U=1$, for 
exponents $\alpha$ that tune through the Anderson-localization transition. 
Unlike the BCS-AA model, the BCS-PRBM does not go through a sharp Anderson-localization transition. 
This can be seen from the wave functions, which show multifractal properties over a wide range of $\alpha$.
In the extended phase, the wave functions with close energies overlap strongly in space and we have a nonzero Chalker correlation function
[Eq.~(5) in the main text]. In the localized phase, the wave functions are localized in small regions and they have little overlap in 
space, resulting in a vanishing Chalker correlation function.
In Figs.~\ref{fig:alphaf-PRBM-U=0.5} and  \ref{fig:alphaf-PRBM-U=1}, 
we show the singularity spectrum of BCS-PRBM model with $U=0.5$ and $U=1$.

\section{Chalker scaling}
For critical single-particle states near an Anderson MIT, the multifractal
enhancement of matrix elements is due to Chalker scaling \cite{SM-Chalker1988,SM-Chalker1990,SM-Cuevas2007,SM-Feigelman2007,SM-Feigelman2010}.
The normalized Chalker correlation function is defined as
\begin{equation}
	C_{E,E_0} 
	= 
	\frac{\int dx \left|\psi_{E_0}(x)\right|^2 \left|\psi_{E}(x)\right|^2}{\frac{1}{2}\left( \int dx \left|\psi_{E_0}(x)\right|^4 + \int dx \left|\psi_{E}(x)\right|^4 \right)}\,,
	\label{eq:Chalker}
\end{equation}
where $\psi_E(x)$ is an exact eigenstate with energy $E$. 
The Chalker scaling correlation characterizes the spatial overlap of eigenstates of a system. 
For $\{E,E_0\}$ close to the MIT, 
one expects $C_{E,E_0} \sim |E - E_0|^{- \gamma}$, where $\gamma = 1 - \tau_2/d$ in $d$ spatial dimensions \cite{SM-Chalker1988,SM-Chalker1990}.
By contrast, $C_{E, E_0}\sim0$ for localized states with nearby energies $\left|E-E_0\right|\ll \delta_l$, where $\delta_l$ the typical level spacing in the localization volume. 
The pairing order parameter contributed by a quasiparticle state $\left|n\right>$ with energy $E_n$ is defined as 
\begin{equation}
	\Delta(E_n) 
	= 
	-
	\frac{U}{N} 
	\sum_i 
	\left<n\right|c_{i\downarrow}c_{i\uparrow}\left|n\right>\,.
  \label{eq:DeltaE}
\end{equation}

\begin{figure}[t]
	\centering
	\includegraphics[width=0.8\textwidth]{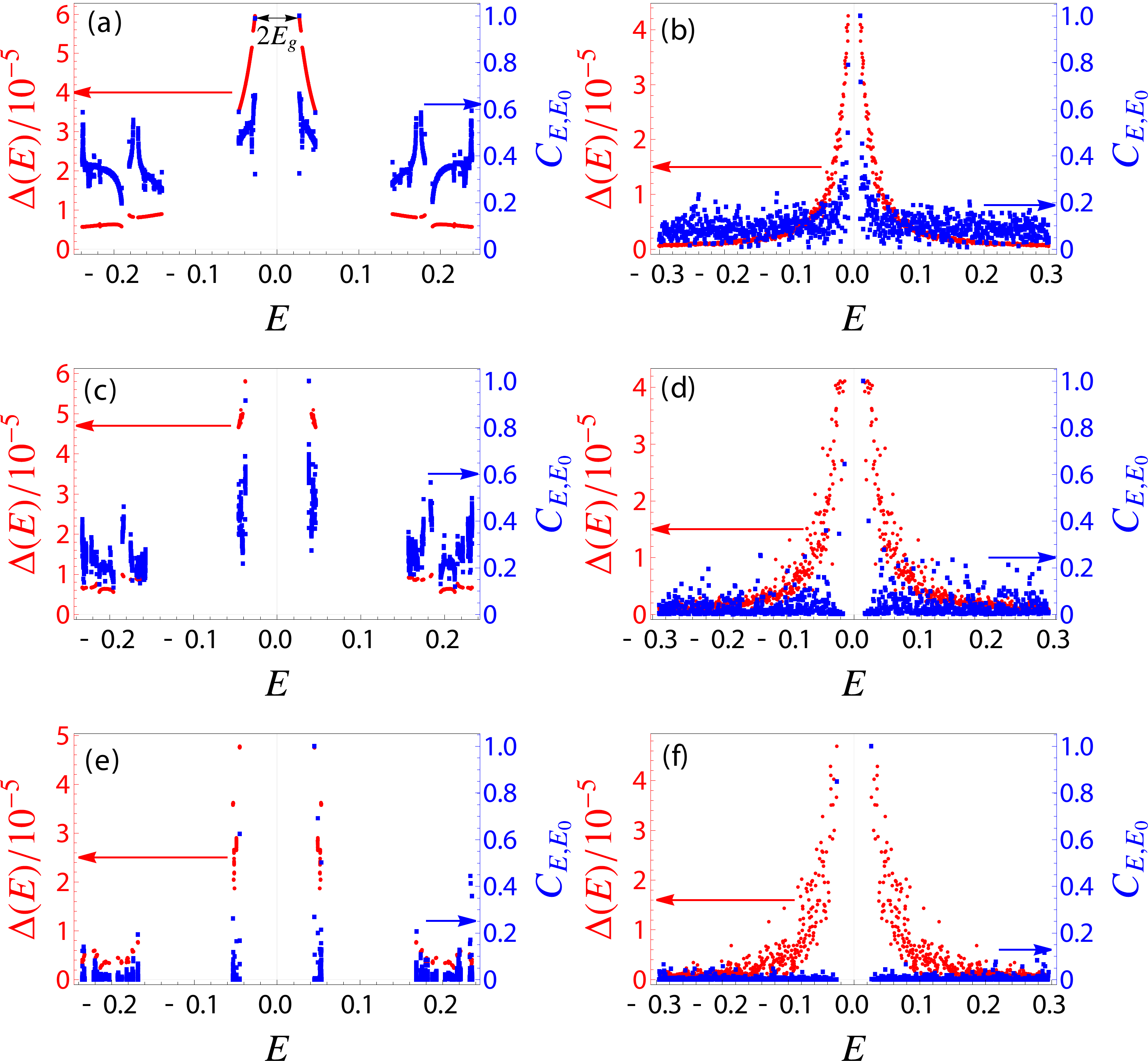}
	\caption{Chalker correlation function $C_{E,E_0}$ [Eq.~(\ref{eq:Chalker})] 
	and $\Delta(E)$ [Eq.~(\ref{eq:DeltaE})] for low-lying quasiparticle states in both the BCS-AA (a,c,e) and BCS-PRBM (b,d,f) models 
	in the extended phase (a,b), 
	localized phase (e,f), 
	and near the critical point (c,d) with $U=0.5$. 
	In the BCS-AA model, we choose $V=1.3$ (a) for the extended phase and $V=1.5$ (e) for the localized phase, 
	while $V = 1.4 \simeq V_c$ (c) at the critical point. 
	In the BCS-PRBM model, $\alpha$ is taken to be $0.9$, $1.0$ and $1.1$ for the extended (b), critical point (d) and localized phase (f), respectively.
	The energy $E_0$ is chosen to correspond to the lowest-lying (positive-energy) quasiparticle state.
	$\Delta(E)$ decays continuously with energy in the extended phase and becomes point-like in the localized phase. 
	In the BCS-AA model, $\Delta(E)$ is almost a constant for states in the same subband at the critical point, 
	and it decays slowly with energy. The Chalker correlator tends towards zero for almost every state $E$ 
	in the localized phase, since the localized wave functions do not overlap in space.
	In the BCS-PRBM model, results are shown for a single realization of the disorder.
}
	\label{fig:ChalkerDeltaE}
\end{figure}

In Fig.~\ref{fig:ChalkerDeltaE}, we show $\Delta(E)$ and the Chalker correlation function $C_{E,E_0}$ near the self-consistently determined MIT, 
with $U=0.5$ and $E_0$ chosen to be the lowest-lying quasiparticle state. In the extended phase of the BCS-AA model 
[Fig.~\ref{fig:ChalkerDeltaE}(a)], $\Delta(E)$ is continuous in the same band and decays rapidly with energy. 
$C_{E,E_0}$ is also (almost) continuous and takes nonzero values. In the localized phase [Fig.~\ref{fig:ChalkerDeltaE}(e)], 
$\Delta(E)$ is point-like and decreases rapidly in energy and $C_{E,E_0}$ takes values close to $0$. 
At the MIT [Fig.~\ref{fig:ChalkerDeltaE}(c)], 
$\Delta(E)$ takes almost constant values in each subband and $C_{E,E_0}$ decreases from the extended phase value and starts to show discontinuous features.

In the BCS-PRBM model [Figs.~\ref{fig:ChalkerDeltaE}(b),(d),(f)], the physics is similar except for the absence of flat bands and Hofstadter gaps. 
$\Delta(E)$ decreases smoothly with energy 
in the extended phase, but exhibits strong scatter in the localized phase. 
Near the critical point, more states are piled around the peak of $\Delta(E)$. $C_{E,E_0}$ is almost zero 
everywhere except for several scattered points when the system is in the localized phase with increasing $\alpha$.